 \documentclass[preprint2]{aastex}

\slugcomment{Accepted by the Astronomical Journal}

\shorttitle{Velocity Dispersion of OB associations}
\shortauthors{H. Kamaya}

\begin{document}


\title{Velocity Dispersion of Dissolving OB Associations  
       Affected by External Pressure of 
       Formation Environment}


\author{ Hideyuki Kamaya }
\affil{Department of Astronomy, School of Science, Kyoto University,
    Kyoto, 606-8502, Japan}

\email{kamaya@kusastro.kyoto-u.ac.jp}



\begin{abstract}
This paper presents a possible way to understand dissolution of OB
associations (or groups).  Assuming rapid escape of parental cloud gas
from associations, we show that the shadow of the formation environment
for associations can be partially imprinted on the velocity dispersion
at their dissolution.  This conclusion is not surprising as long as
associations are formed in a multiphase interstellar medium, because the
external pressure should suppress expansion caused by the internal
motion of the parental clouds.  Our model predicts a few km s$^{-1}$ as
the internal velocity dispersion. Observationally, the internal velocity
dispersion is $\sim 1$ km s$^{-1}$ which is smaller than our prediction.
This suggests that the dissipation of internal energy happens before the
formation of OB associations.

\end{abstract}




\section{Introduction}

Open clusters and OB associations (or groups) 
have been investigated from various points of view, 
considering spectral types of member stars (e.g.,  
Sanford 1949; Svolopoulos 1961; Levato \& Malaroda 1975;
Abt \& Levato 1975; Trumpler 1988; Morrell, Garcia, Levato 1988; 
Sears \& Sowell 1997; Wang \& Hu 2000; Piatti et al. 2002),
mass function (e.g.
Jaschek \& Jaschek 1957; Frolov 1975; Lee \& Kim 1983;
Ann \& Lee 1989; Phelps \& Janes 1993; de La Fuente Marcos 1995;
von Hippel et al. 1996; Sagar \& Griffinths 1998; 
Barrado y Navascues et al. 2001; Prisinzano et al. 2003),
abundance (e.g.
Demarque \& Heasley 1971; Chaffee, Carbon, Strom 1971;
Zappala 1972; McClure 1972; Barry \& Cromwell 1974;
Norris \& Hawarden 1978; Claria 1979; Panagia \& Tosi 1980;
Cameron 1985a, 1985b; Smith \& Suntzef 1987; Gilroy 1989;
Boesgaard 1991; Gratton \& Contarini 1994; Edvardsson et al. 1995;  
Tiede, Martini, Frogel 1997; 
Sarajedini 1999; Gonzalez \& Wallerstein 2000; 
Randich et al. 2001; Mathys et al. 2002; 
Friel et al. 2002), 
kinematics(e.g.
Gieseking 1981; Hron 1987; Lynga \& Palous; Friel 1988; 
Sagar \& Bhatt 1989; Scott, Friel, \& Janes 1995;
Sanner et al. 2000; Gonzalez \& Lapasset 2000), and so on.

The origin and evolution of clusters are also a classical set
of problems 
(e.g. Starikova 1966; Kaliberda 1973; 
Palous et al., 1977; Burki 1978; van den Bergh 1981; Turner 1985;
Danilov 1987; Battinelli \& Capuzzo-Dolcetta 1991;
Phelps \& Janes 1994; van den Ancker et al. 1997;
Elmegreen \& Efremov 1997; Belikov et al. 2000;
Garcia \& Mermilliod 2001).
Recently, understanding  these problems has been recognized
to be very important if we want
to know how disks of spiral galaxies formed and evolved 
(e.g., Efremov \& Elmegreen 1998a, b;
Nomura \& Kamaya 2001),
because star complexes and associations are fundamental
and elementary cells of star formation (e.g. Efremov 1995).

This paper tries to investigate some elemental connections between
the formation and evolution of clusters. In particuler, 
the dissolution process of
OB associations (or groups) is studied.

\section{Analysis}
First of all, we define a {\it parental cloud} in which OB association
is formed. The parental cloud is made of the typical cold component of
interstellar medium (ISM) of the Milky Way. We focus on  
the origin and evolution of OB associations at the present epoch.

\subsection{Assumptions}
We set up our problem with the following four assumptions:
(1) At the formation epoch, parental clouds are in virial equilibrium,
affected by the external pressure of the ISM.  
This initial condition is necessary
since the various phases of ISM are roughly in  pressure equilibrium 
(Myers 1978).  Pressure jumps at surfaces of parent clouds are 
unexpected or of little significance.
(2) Gas of parent clouds is removed rapidly.  This means there is a
sufficient chance for parental clouds or cloudlets to be in pressure
equilibrium before star formation starts inside them.  
In other words, the initial
temperature and density of gas content of parental clouds are 100 K and
100 cm$^{-3}$, respectively,
with an external pressure of $\sim 10^{-12}$ ergs cm$^{-3}$ .
This becomes possible if size of parental clouds is initially small.
After growing via coagulation of blobs of cold phase ISM
and/or accumulation of warm phase ISM due to radiative cooling,
dynamical collapse occurs when size of parental clouds becomes larger
than the Jeans-length.  Soon after the formation of clusters, the
parental gas is removed very rapidly by  the activity of massive
stars. Thus, as long as some massive stars are formed in the multi-phase
ISM, rapid removal assumption for the parental clouds can be very
reasonable.  
(3) To show the importance of the external pressure, we drop the effect
of angular momentum and the figure rotation of the parental cloud.  
(4) To find the final velocity dispersion, we adopt energy conservation
law for the stellar component.

\subsection{Virial equilibrium }

The virial equilibrium for a parental cloud is expressed as
\begin{eqnarray}
2K_0 + W_0 = 3 p_0 V_0  , 
\end{eqnarray}
where $V$ is volume of the parental cloud,
 $K$ is kinetic energy, $p$ is external pressure,
$W$ is gravitational energy, and the subscript  $0$ denotes the initial
condition (e.g. Hill 1980). 
Denoting velocity dispersion as $<v_0^2>$, we estimate
\begin{eqnarray}
<v_0^2> = \frac{GM_0}{2R_0} + \frac{3p_0 V_0}{M_0}.
\end{eqnarray}
Here, $G$ is the gravitational constant, $M$ is mass
of a parental cloud, $R$ is characteristic size of a parental cloud,
and the subscript $0$ again  denotes the initial condition.

When stars are formed in a parental cloud, the total energy of stellar
component is estimated as 
\begin{eqnarray}
E_{\rm star} = K_{\rm star} + W_{\rm star}
= \frac{M_{\rm star} <v_0^2>}{2}
 -\frac{G M_{\rm star}^2}{2 R_0}
\end{eqnarray}
where $E_{\rm star}$ is total energy,
$K_{\rm star}$ is kinetic energy, $W_{\rm star}$ is gravitational
energy, and $M_{\rm star}$ is total mass of stellar component.
Here, rapid removal assumption is adopted.

\subsection{Energy conservation}

As well-known, OB associations are gravitationally unbound system.
Thus, the sign of the total energy is always positive. We let $<v_1^2>$ be
the current velocity dispersion, and 
estimate the energy for the dissolving association as
\begin{eqnarray}
E_1 = \frac{M_{\rm star} <v_1^2>}{2}
\end{eqnarray}
where $E_1$ is total energy of the association.
At this epoch, 
the gravitational energy of the association has become negligible.

With energy conservation,
\begin{eqnarray}
E_1 = E_{\rm star}, 
\end{eqnarray}
we find
\begin{eqnarray}
<v_1^2> = 
\frac{GM_{\rm star}}{2R_0} \left( \frac{M_0}{M_{\rm star}} - 2\right)
+ \frac{3p_0V_0}{M_0} . 
\end{eqnarray}
It should be noted that we are considering gravitationally unbound system.
That is, the star formation efficiency in mass, 
$\Gamma_{\rm SFE} \equiv M_{\rm star} / M_0$, is smaller than $\sim 0.5$
(i.e. $M_0/M_{\rm star} > 2.0$).

\subsection{Estimate of velocity dispersion}

The two terms of the right hand side of Eq.(6) are denoted as
\begin{eqnarray}
A_1 = 
\frac{GM_0}{2R_0} \left( 1 - 2 \Gamma_{\rm SFE} \right)
\end{eqnarray}
and
\begin{eqnarray}
B_1 = \frac{3p_0V_0}{M_0}, 
\end{eqnarray}
respectively.

For $A_1$, it is reasonable for us to estimate $R_0 \sim R_{\rm J}$,
where $R_{\rm J}$ is half of the Jeans wavelength: 
\begin{eqnarray}
R_{\rm J} = \frac{a_0}{2} \left( \frac{\pi}{G\rho_0} \right)^{0.5}
\end{eqnarray}
where $a_0$ is the sound speed and $\rho_0$ is mean density
at the formation epoch.
With hypothesis of spherical geometry for a parental cloud,
we find 
\begin{eqnarray}
A_1 = \frac{\pi ^2}{6} \left( 1 - 2 \Gamma_{\rm SFE} \right)
      \times a_0^2 .
\end{eqnarray}

The other term $B_1$ is also expressed in terms of $a_0$;
\begin{eqnarray}
B_1 =  3 a_0^2
\end{eqnarray}
with the assumption of spherical geometry for a parental cloud.
Hence, we find that both $A_1$ and $B_1$ are 
on the order of $a_0^2$.
In the current case, $a_0$ is about 1 km s$^{-1}$. 
Hence, $\sqrt{<v_1^2>}$ is on the order of 1 km s$^{-1}$ as
\begin{eqnarray}
\sqrt{ <v_1^2> }= 
\left[\frac{\pi ^2}{6} \left( 1 - 2 \Gamma_{\rm SFE} \right) 
      + 3 \right]^{0.5} a_0 .
\end{eqnarray}
Note that $B_1$ (i.e. contribution from the external
pressure) can dominate $A_1$, if $\Gamma_{\rm SFE} \sim 0.1$ and/or the
geometry of parental clouds is not far from round shape.

\section{Discussion}
\subsection{Aspect ratio of cloud configuration}
In general, geometrical configuration of interstellar clouds is not
spherical. It seems to be filament.  If angular momentum and figure
rotation of the clouds are neglected, the virial equilibrium for a
filamentally cloud is expressed roughly as
\begin{eqnarray}
2K_0 + W_0 =  4 \pi R^2_0 Z_0 p_0 , 
\end{eqnarray}
where $R_0$ is the size of semiminor axes and $Z_0$ is that of semimajor
axes.  {}From this equation, we find that as long as the aspect ratio,
$R_0/Z_0$, of the cloud configuration is not much smaller than unity,
our spherical assumption is not so crude.  Indeed, the aspect ratio of
the parental clouds is not so small, since the self-gravity of the
parental cloud is comparable to the internal pressure at the formation
epoch.  We remember $R_0 = R_{\rm Jeans}$ which means that the internal
thermal energy is comparable to the self-gravitational energy.  Then, we
confirm qualitatively that $B_1$ can dominate or be comparable to $A_1$
for determining the final velocity dispersion of associations.

For general purpose, the scalar virial theorem is not adopted.  The
tensor virial theorem (Weber 1976) is necessary, especially when the
angular momentum and figure rotation are important.  If the effects of
angular momentum and figure rotation are comparable to that of the
external pressure, the coupling among the three effects (external
pressure, angular momentum, and figure rotation) may be reflected in the
internal velocity dispersion of OB associations. In future work, we try
to examine the angular momentum and figure rotation rigorously.

If the overall stellar distribution of associations is far from
spherical, our consideration may become meaningless.  In Efremov (1995),
spatial distribution of Cepheids and membership of OB association are
examined. According to his paper, fortunately, the configuration of OB
associations is not so far from spherical (see also Parker et al. 2001).
Furthermore, stellar density distribution of $\eta$ Cha cluster member
seems to be spherical (Mamajek, Lawson, \& Feigelson 2000).

\subsection{Gravitational effect of disk}
In the final process of dissolution of OB associations, the external
gravity of the Milky Way can be effective. However, for
young and/or intermediate age OB associations, the external gravity field 
is not so critical (e.g. Brown, Dekker, \& de Zeeuw 1997).  This
is because, the initial size of associations can be less than $\sim 10$
pc (Brown et al. 1997), while the length scale for the tidal force from
the galactic disk can be about 10 pc (e.g. Keenan, Innanen, \& House
1973).  Thus, velocity dispersion of young associations can retain
the information of parental clouds affected by the external pressure.

\subsection{Comparison to observation}
OB associations expand because of their unbound nature.  The expansion can
be detected if and only if the correct mean streaming motion of the
association with respect to the Sun is subtracted from the observed
proper motion and radial velocities (e.g. Steffey 1973).  Even if we
use the data of {\it Hipparcos}, this is a very difficult task (de
Zeeuw et al. 1999). Fortunately, de Bruijne (1999) has succeeded in 
finding the internal velocity dispersion of nearby OB associations 
with the assumptions 
of isotropy and compactness.  According to this work, the internal
velocity dispersion of $\sqrt{ <v_{\rm obs}^2> }$ is about $ 1.0 - 1.5 $
km s$^{-1}$.  Our theoretical model predicts 
$\sqrt{ <v_1^2> } \simeq 2$ km s$^{-1}$ if $\Gamma_{\rm SFE} = 0.1$.
The author thinks this estimate is consistent with observational
constraint.

However, our estimate of $<v_{\rm
1}^2> \simeq 4$ km$^{2}$ s$^{-2}$ is slightly larger than 
$<v_{\rm obs}^2>$ $\sim$ 1 km$^{2}$ s$^{-2}$, which suggests
two points.  (1) The sound speed of the parental clouds is reduced by
the formation of hydrogen molecules, which increases the mean molecular
weight.  This effect decreases $<v_{\rm 1}^2>$ by a factor of 0.5, that
is $<v_{\rm 1}^2> \to 2$ km s$^{2}$ s$^{-2}$.  
The difference between $<v_{\rm 1}^2>$ and $<v_{\rm obs}^2>$ becomes
small, so further dissipation of internal energy of the
parental clouds may be needed to reduce $<v_{\rm 1}^2>$.  
(2) Dissipation of energy
during the formation of OB associations can be important. This decreases
the internal velocity dispersion before the rapid gas removal.  If so,
the observational internal velocity of $\sqrt{ <v_{\rm obs}^2>}$ should
be always smaller than our $\sqrt{<v_{\rm 1}^2>}$.  Our theory for 
$\sqrt{<v_{\rm 1}^2>}$ presents the upper-bound for internal velocity
dispersion of young OB associations.

We would like to comment on some detailed studies of the velocity
dispersion of open clusters. McNamara \& Sekiguchi (1986) reported 
$\sim 1.0$ km s$^{-1}$ for M 35.  Girard et al. (1989) insist velocity
dispersion of M67 is $\sim 0.8$ km s$^{-1}$ (see also McNamara \&
Sanders 1978).

\subsection{Comments on interstellar turbulence}
Throughout this paper, we have mentioned on the effect of the
interstellar turbulence. Here, some comments are presented.  To
determine $a_0$, it is natural for us to regard $a_0^2$ as the sum of two
components, thermal and turbulent motion, which suggests that the 
kinematics of young associations reflects the turbulent motion of the
parental clouds (Elmegreen, Kimura, \& Tosa 1995).  This last possibility
may explain one of the elemental processes of the relation between age
and distance among clusters and associations (Efremov \& Elmegreen
1998a, 1998b).

\subsection{Comments on numerical work}
The dissolution process is examined precisely by Brown, Dekker, \& de Zeeuw
(1997). The internal velocity dispersion of their initial conditions is
a few km s$^{-1}$.  Our model favors a similar internal velocity.
According to their results, the discrepancy between
the kinematic and nuclear ages of OB associations is attributed to
underestimates of the kinematic age.  This may be occurred because of
overestimation of the internal velocity dispersion, as in our model.
In the author's opinion, the dissipation of thermal and
kinetic energy of the parental clouds before the formation of OB
associations is important in decreasing the internal (i.e. expanding)
velocity of associations.

The importance of the external pressure is also suggested numerically by
Elmegreen, Kimura \& Tosa (1995). According to their numerical work, the
internal velocity dispersion of forming OB association is determined by
the shock condition. Obviously, the thermal condition of the parental
clouds is determined by the external pressure via flux-conservation laws
of mass, momentum, and energy.  The difference between their work and
ours is very clear.  That is, we insist that even
if the associations are isolated at their formation epoch (i.e. not
sequential formation mode), the external pressure can affect their
internal kinematics.  The isolated formation mode of OB associations is
important when the turbulent speed of the ISM is larger than the expansion
speed of the H{\sc II} region (Nomura \& Kamaya 2001).  Of course, further
theoretical consideration is necessary.

\section{Summary}
We show that the internal velocity dispersion at the dissolution epoch
of OB associations is affected by the external pressure on their
parental clouds.  
The observations show 
the velocity dispersion of dissolving association to be $\sim 1$ km s$^{-1}$,
which is smaller than our estimate of a few km s$^{-1}$.
The conclusion is that motion of forming stars in OB associations 
may have been decreased by energy dissipation inside the parental cloud.

\acknowledgments
I am thankful to the referee for his/her advice and
excellent refereeing, which improved my paper very much in both
content and presentation.
I am also grateful to Prof.S.Mineshige and Dr.A.K.Inoue for their 
encouragement.
This work is supported by the Grant-in-Aid for the 21st Century 
COE {\it Center for Diversity and Universality in Physics} from 
the Ministry of Education, Culture, Sports, Science and Technology 
of Japan.

\end{document}